# Anomalous transparency of water-air interface for low-frequency sound

*Oleg A. Godin*

Cooperative Institute for Research in Environmental Sciences, University of Colorado and NOAA/Earth System Research Laboratory, Boulder, CO 80305. E-mail: oleg.godin@noaa.gov

**Abstract.** Sound transmission through water-air interface is normally weak because of a strong mass density contrast. Here we show that the transparency of the interface increases dramatically at low frequencies. Rather counterintuitively, almost all acoustic energy emitted by a sufficiently shallow monopole source under water is predicted to be radiated into atmosphere. Physically, increased transparency at lower frequencies is due to the increasing role of inhomogeneous waves and a destructive interference of direct and surface-reflected waves under water. The phenomenon of anomalous transparency has significant implications for acoustic communication across the water-air interface, generation of ambient noise, and detection of underwater explosions.

Because of the stark mass density contrast between air and water, water-air interface is normally considered as a perfectly reflecting, pressure-release boundary in underwater acoustics or a rigid boundary in atmospheric acoustics [1]. Ray-theoretical calculations predict weak coupling between sound fields in air and water, with an energy transmission coefficient on the order of the ratio of acoustic impedances of air and water [2, 3]. However, at infrasonic frequencies, underwater sources are typically located within a fraction of the wavelength from the interface, and ray calculations cease to be applicable. Here we show that the water-air interface is anomalously transparent for sound radiated by shallow sources, and almost all the energy emitted under water can be radiated into the air. For a monopole source at a depth which is small compared to acoustic wavelength, the ratio of energy radiated into the air to the total emitted energy is larger by a factor up to 3,400 under normal conditions than it is for the same source located a wavelength or more from the interface. We also show that the increase in transparency of the interface and the absolute value of the acoustic power radiated into the air are sensitive to the type of the underwater acoustic source.

Instead of being an almost perfect mirror, as previously believed, water-air interface can be a good conduit of low-frequency underwater sound into the atmosphere. The anomalous transparency of the water-air interface may have significant implications in problems that range from generation of low-frequency ambient noise in the air by bubbles collapsing under water and heating of the upper atmosphere due to absorption of infrasound to understanding the role of hearing in avian predation of aquatic animals and acoustic monitoring and detection of powerful underwater explosions for the purposes of the Comprehensive Nuclear-Test-Ban Treaty.

Previous theoretical [4-13] and experimental [3, 7, 9, 10, 12, 14-17] studies of sound transmission through the air-water interface concentrated on acoustic fields under water due to airborne sources, primarily because of the existence of powerful noise sources in the atmosphere (such as helicopters [16], propeller-driven aircraft [10, 13, 15], and supersonic transport with their attending sonic booms [9, 11, 12, 17]) and possible effects of these man-made sources on marine life [9, 16].

We apply a theory of acoustic fields in layered media [3, 18] to develop a full-wave description of sound transmission from water to air. Penetration of sound through the interface will be characterized by acoustic transparency, defined as the ratio of acoustic power radiated into air to the total acoustic power emitted by an underwater continuous wave source. Introduce Cartesian coordinate system $\mathbf{r} = (x, y, 0)$ with a vertical coordinate $z$ increasing downward (Fig. 1A). Plane $z = 0$ separates homogeneous half-spaces $z > 0$ (water) and $z < 0$ (air), where sound speeds and mass densities are $c_1, \rho_1$ and $c_2 = c_1/n, \rho_2 = m \rho_1$, respectively. Representative values of the refraction index $n$ and mass density ratio are $n = 4.5, m = 1.3 \cdot 10^{-3}$. It is the smallness of $m$ and $n^{-2}$ compared to unity that is responsible for peculiarities of sound transmission through the water/air interface.

Let a point source be situated at a point $(0, 0, z_0)$, $z_0 > 0$ in water. Acoustic pressure $p_1$ due to the source in the absence of an interface, wave $p_2$ reflected from the interface into water, and wave $p_3$ refracted into air are given by integrals over plane waves [18]:

$$p_j(\mathbf{r}) = (2\pi)^{-1} i \int d^2\mathbf{q} \, e^{i\mathbf{q}\cdot\mathbf{r} + iv_1 z_0} v_1^{-1} Q_j(\mathbf{q}), \quad j = 1, 2, 3, \qquad (1)$$

$$Q_1 = S_1(\mathbf{q}) e^{i v_1 (z - 2 z_0)}, z > z_0; \quad Q_1 = S_2(\mathbf{q}) e^{-i v_1 z}, z < z_0, \qquad (2)$$

$$Q_2 = S_2(\mathbf{q}) V(q) e^{i v_1 z}, z > 0; \quad Q_3 = S_2(\mathbf{q}) W(q) e^{-i v_2 z}, z < 0, \qquad (3)$$

where $\mathbf{q} = (q_1, q_2, 0)$, $q \equiv |\mathbf{q}|$; $v_s = (k_s^2 - q^2)^{1/2}$, $\mathrm{Im}\, v_s \geq 0$; $k_s = \omega / c_s$, $s = 1, 2$; $\omega$ is sound frequency, and

$$V = (m v_1 - v_2)/(m v_1 + v_2), \quad W = 2 m v_1 / (m v_1 + v_2) \qquad (4)$$

are Fresnel reflection and transmission coefficients [3] for an incident plane wave with the wave vector $(q_1, q_2, v_1)$. Functions $S_1(\mathbf{q})$ and $S_2(\mathbf{q})$ are plane-wave spectra of the field emitted by the source downward and upward, respectively. These functions determine the source type. In particular, if $S_1 = S_2 = 1$, we have a monopole sound source with $p_1 = p_0$, $p_0 = R^{-1} \exp(i k_1 R)$, $R = \left[ x^2 + y^2 + (z - z_0)^2 \right]^{1/2}$ [18]. When $S_1 = -S_2 = i v_1 / k_1$, we have a vertically oriented dipole source with $p_1 = k_1^{-1} \partial p_0 / \partial z$. The spectra $S_1 = S_2 = i q_1 / k_1$ correspond to a horizontal dipole source with $p_1 = k_1^{-1} \partial p_0 / \partial x$.

Wave vectors of reflected and refracted plane waves are $(q_1, q_2, v_1)$ and $(q_1, q_2, -v_2)$. According to Snell's law [3], the horizontal components of a wave vector do not change at reflection and refraction. Plane waves with $0 \leq q \leq k_1$ are homogeneous in water (i.e., $\mathrm{Im}\, v_1 = 0$) and give homogeneous refracted waves in air with refraction angles $0 \leq \theta_2 \leq \delta$, $\delta \equiv \arcsin n^{-1}$. Plane waves with $k_1 < q \leq k_2$, which are inhomogeneous (evanescent) in water (i.e., $\mathrm{Im}\, v_1 > 0$), give homogeneous refracted waves in air with refraction angles $\delta < \theta_2 \leq \pi/2$ (Fig. 1A). When $q > k_2$, both incident and refracted waves are evanescent. Under normal conditions, the critical angle $\delta \approx 13°$.

Acoustic power flux $J_a$ into air can be calculated by integrating the normal component of the acoustic power flux density $(2\omega\rho)^{-1} \text{Im}(p^* \nabla p)$ [3] over the interface $z = 0$. Here, the asterisk denotes complex conjugation. Using eqs. (1)-(4), we obtain

$$J_a = \frac{J_0}{4\pi k_1} \int_{q<k_2} d^2\mathbf{q} \, |S_2(\mathbf{q})|^2 \exp(-2z_0 \text{Im}\,\nu_1) \text{Re}\left(\frac{1-|V|^2 + 2i\,\text{Im}\,V}{\nu_1}\right), \quad (5)$$

where $J_0 = 2\pi/\rho_1 c_1$ is the acoustic power radiated by the waterborne monopole source in the absence of the interface. The dipole sources defined above radiate acoustic power $J_d = J_0/3$ in unbounded water.

The total power output of the generic source $J_t = J_a + J_w$ includes the acoustic power flux [19]

$$J_w = \frac{J_0}{4\pi k_1} \int_{q<k_1} \frac{d^2\mathbf{q}}{\nu_1} \left|S_1(\mathbf{q}) + V(q) S_2(\mathbf{q}) e^{2i\nu_1 z_0}\right|^2, \quad (6)$$

which is carried to infinity within water [20]. With eqs. (5) and (6), the power fluxes $J_a$ and $J_w$ can be readily calculated numerically or evaluated analytically using the small parameters $m$ and $n^{-2}$ of the problem [19].

When $m \ll 1$, $V \approx -1$ except for plane waves with $k_2 - q = O(m^2)$. The power output of sound sources (Fig. 1B) is very close to that in the case of a pressure-release boundary, where $V = -1$, with a possible exception for source depths $z_0$, which are small compared to the acoustic wavelength in water. For sources with anti-symmetric plane-wave spectra ($S_1 = -S_2$) power output $J_t$ nearly doubles at $z_0 \to 0$ compared to its value at large $z_0$, because of the constructive interference of incident ($p_1$) and reflected ($p_2$) waves (Fig. 1B). For sources with symmetrical spectra ($S_1 = S_2$) near the pressure-release boundary, the power output vanishes at $z_0 \to 0$ because of the destructive interference of incident and reflected waves. When $0 < m \ll 1$, $J_t$ remains finite for all source depths and has a deep minimum (Figs. 1C and 1D).

Directivity of radiation in the air is characterized by angular density $D$ of the acoustic power flux: $J_a = \int_0^{\pi/2} D(\theta_2) d\theta_2$. The angular density is shown in Figs. 2A-C. For shallow sources, the bulk of the radiation occurs at refraction angles $\theta_2 > \delta$. Radiation in such directions rapidly decreases with the depth increase, because of a decrease in amplitude of incident evanescent waves at the interface $z = 0$, as described by the exponential factor in eq. (5). Ray theory accounts only for radiation in the directions $\theta_2 \leq \delta$. This radiation is due to incident homogeneous plane waves and is independent of the source depth.

The relative contribution of inhomogeneous waves to sound transmission through the interface is also sensitive to refraction index and source type (Figs. 3A - C) but is insensitive to the density ratio $m$ as long as the latter is small. Inhomogeneous waves play a dominant role when acoustic frequency is low and refraction index is large. According to eqs. (4) and (5), for the water-air interface, the contribution of inhomogeneous waves exceeds that of homogeneous waves by the factors $2n^2 \left[1 + O(n^{-2})\right] \approx 40$ and $(8n^4/3)$ $\times \left[1+O\left(n^{-2}\right)\right] \approx 1100$ for very shallow monopole and dipole sources, respectively. As long as the contribution of homogeneous waves is independent of source depth and frequency, these factors also determine the increase in the absolute value of the acoustic power transmitted into air when a source depth and/or frequency decrease, so that that the nondimensional source depth $k_1 z_0$ changes from $k_1 z_0 \gg 1$ to $k_1 z_0 \ll 1$. The role of inhomogeneous waves and, consequently, the increase in sound transmission into air are even greater for higher-order multipole sources due to the greater weight of plane waves with $1 < q < n$ in their spectra.

Because of the contribution of inhomogeneous waves, the acoustic transparency $J_a/J_t$ of the water-air interface rapidly grows with diminishing source depth, when $k_1 z_0 < 1,$ from its ray-theoretical value $O(m/n)$ at $k_1 z_0 \gg 1$ (Figs. 4A and 4B). For sound sources with symmetrical spectra ($S_1 = S_2$), the transparency closely approaches unity: $J_w = J_a\, O(m)$ at $z_0 \to 0$, i.e., almost all emitted energy is radiated into the air (Fig. 4C). Although counterintuitive, this phenomenon is easy to understand. Indeed, when such a source is on the interface, acoustic pressure in both air ($p_3$) and water ($p_1 + p_2$) is proportional to the small parameter $m$ (see eqs. (1) – (4)). Then, because of the different mass densities of the two media, the acoustic power flux in air is $J_0\, O(m)$, while the acoustic power flux in water is $J_0\, O(m^2)$.

In summary, contrary to the conventional wisdom based on ray-theoretical predictions and observations at higher frequencies, infrasonic energy from localized waterborne sources can be effectively transmitted into air. We have demonstrated theoretically that water-air interface is anomalously transparent to low-frequency acoustic waves. The phenomenon of anomalous transparency occurs when a sound source is located at a shallow depth, meaning that the depth is a fraction of acoustic wavelength. For shallow sources, acoustic intensity in the air increases dramatically due to energy transfer from the source by evanescent waves under water. Almost all emitted acoustic energy is channeled into the air when sound is generated by a shallow monopole source or any other localized source with a radiation pattern symmetric with respect to the horizontal plane. For such sources, an increased power flux into the air due to evanescent waves is accompanied by a decrease in downward acoustic power flux due to the destructive interference of direct and surface-reflected waves under water.

19. For details of analysis, see Appendix.

20. Actual water-air interface is a rough surface rather than a plane. For low-frequency sound, roughness elevation $h$ is small compared to other relevant spatial scales. Assuming $k_2 h \ll 1$, $h/z_0 \ll 1$, the effect of surface roughness on sound transmission and sound source output proves to be negligible, see Appendix.

21. This work was supported, in part, by the US Office of Naval Research. We thank M. Charnotskii and I. M. Fuks for discussions.10

**Figure Captions**

**Fig. 1.** Schematic of sound reflection and refraction at a plane water-air interface (A) and power output of monopole (red), horizontal dipole (green), and vertical dipole (blue) sound sources (B-D). The power output $J_t$ is normalized by its value in the absence of the interface; $k_1$ and $z_0$ are the acoustic wavenumber in water and source depth. A set of horizontal lines in panel A represents an evanescent plane wave emitted by the source, with length of the lines corresponding to the wave amplitude. The power output of a source under water-air interface (panels C and D) is noticeably different from the power output of the same source under a pressure-release boundary (panel B) only when non-dimensional source depth $k_1 z_0 \ll 1$.

**Fig. 2.** Directivity $D$ of sound radiated into air by a waterborne monopole (A), horizontal (B), and vertical (C) dipole sources. Refraction index and mass density ratio are $n = 4.5$, $m = 1.3 \cdot 10^{-3}$. Non-dimensional source depth $k_1 z_0 = 0.1$ (1), 0.2 (2), 0.4 (3), 0.5 (4), 0.6 (5), 0.8 (6), and 1.0 (7).

**Fig. 3.** Significance of inhomogeneous waves in sound transmission through water-air interface. Ratio $R$ of acoustic power fluxes into air due to inhomogeneous (evanescent) and homogeneous incident plane waves is shown as a function of non-dimensional source depth $k_1 z_0$ for monopole (A), horizontal dipole (B), and vertical dipole (C) sources located under an interface with strong density contrast ($m \ll 1$) and three values of the refraction index: $n = 1.1$ (blue), 1.5 (green), and 4.5 (red).

**Fig. 4.** Acoustic transparency of water-air interface as a function of source depth for waterborne monopole (red), horizontal (green), or vertical (blue) dipole sources (A), with details of the depth-dependence for deeper (B), and shallower (C) sources. Refraction index and mass density ratio are $n = 4.5, m = 1.3 \cdot 10^{-3}$.

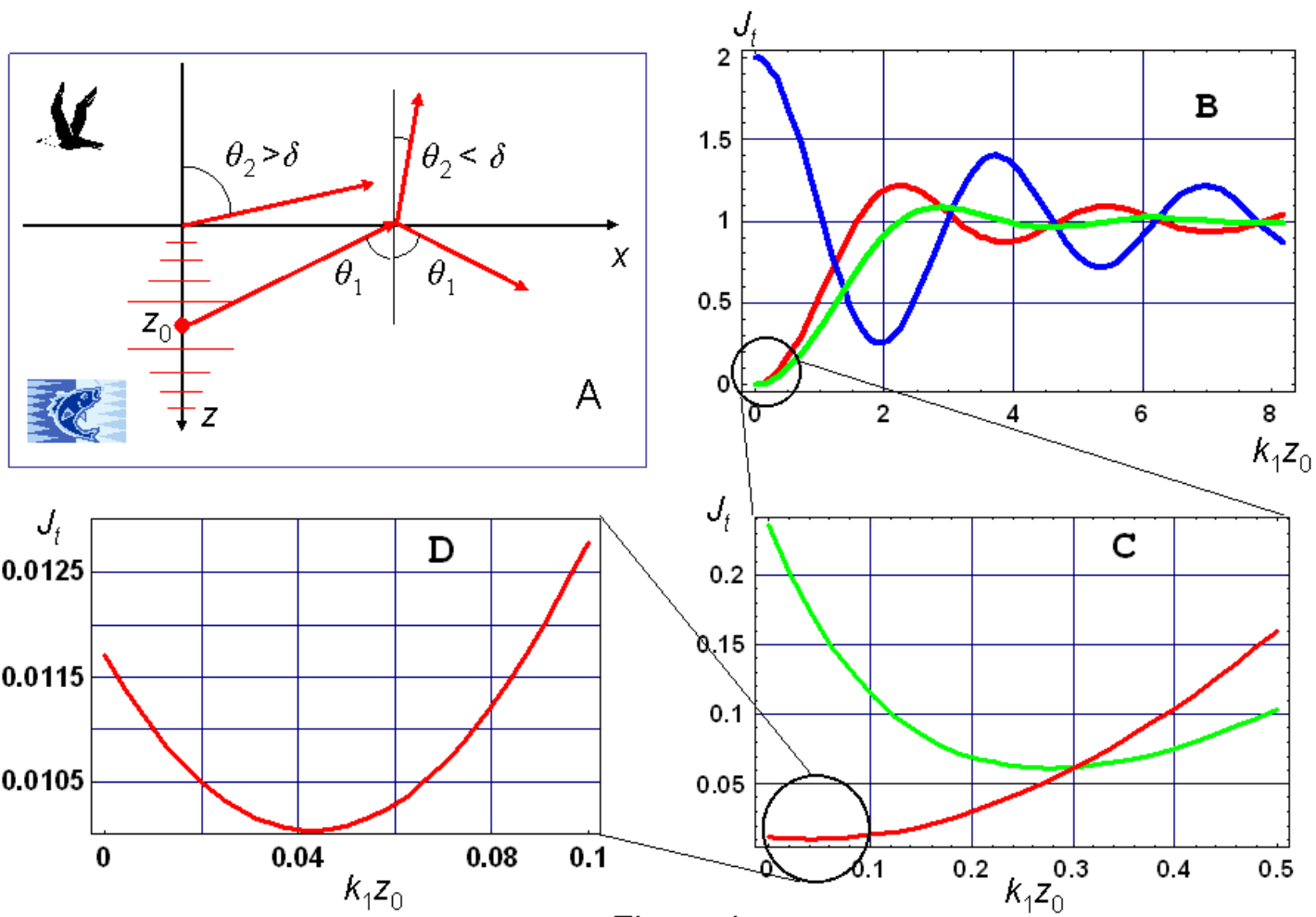

Figure 1

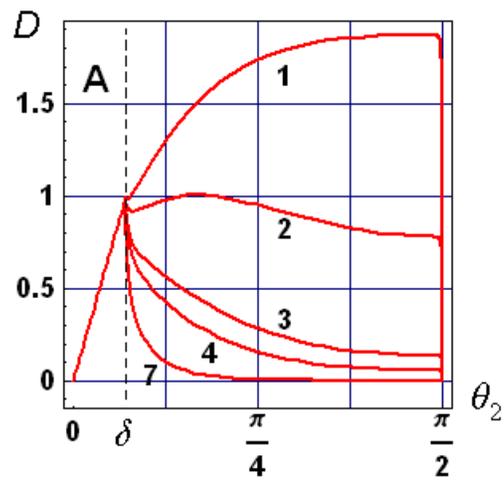
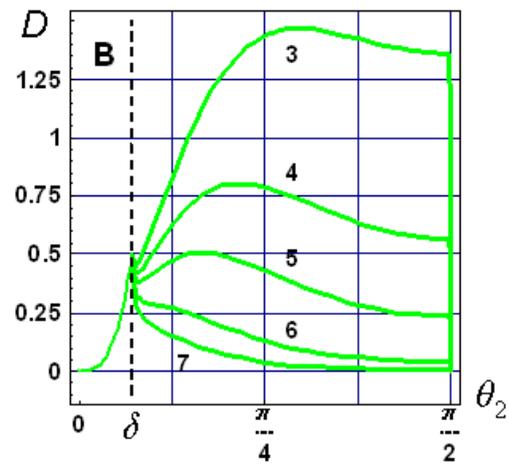
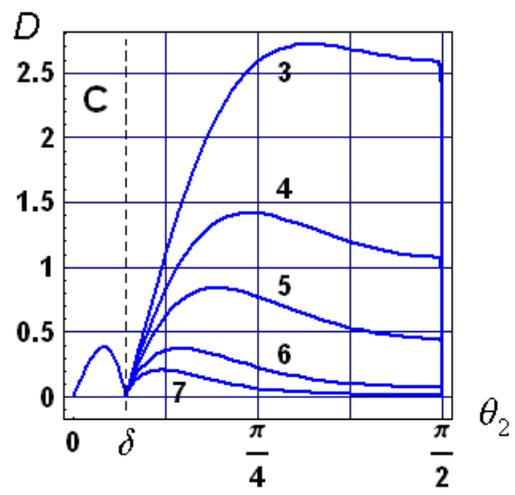

Figure 2

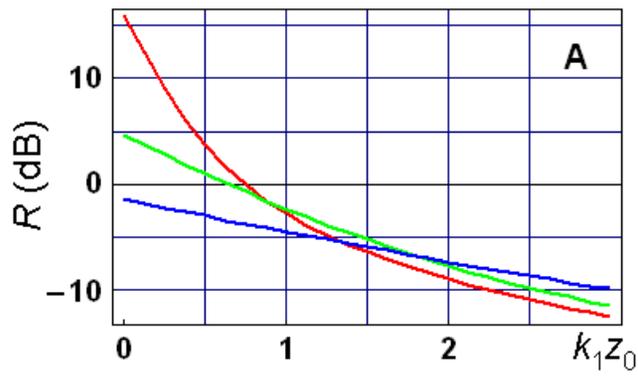
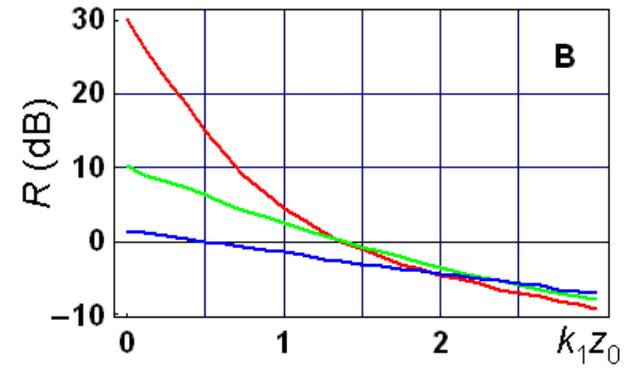
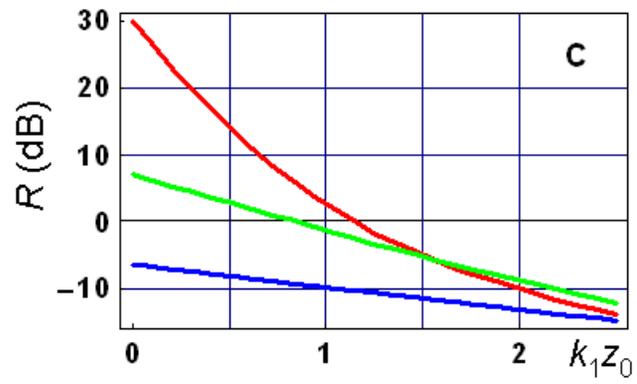

Figure 3

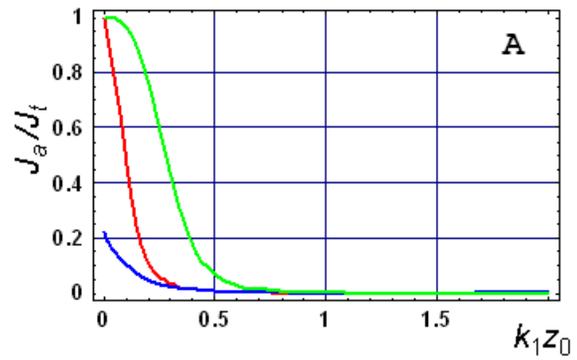
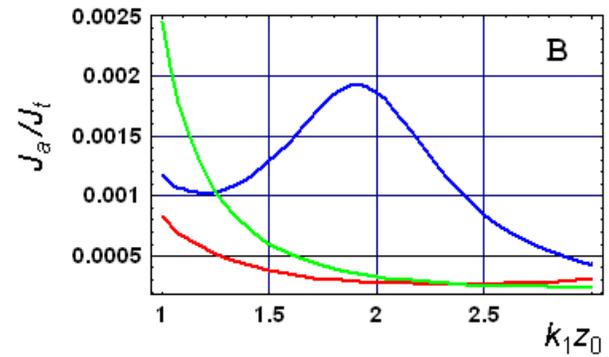
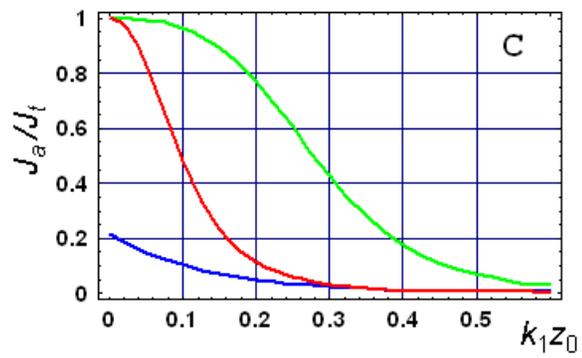

Figure 4

## Appendix

### *Quantifying the acoustic transparency of a plane interface*

Energy flux into the air should be compared with the total acoustic energy flux emitted by the source. In a homogeneous fluid, from the general relation $(2\omega\rho)^{-1}\operatorname{Im}(p^*\nabla p)$ [A1] between acoustic power flux density and acoustic pressure and explicit expressions $p_1 = p_0$, $p_0 = R^{-1}\exp(ik_1 R)$, $R = \left[x^2 + y^2 + (z-z_0)^2\right]^{1/2}$ [A2] (for the monopole source), $p_1 = k_1^{-1}\partial p_0/\partial z$ (for the vertical dipole), $p_1 = k_1^{-1}\partial p_0/\partial x$ (for the horizontal dipole), one easily finds that the total radiated power flux is $J_0 = 2\pi/\rho_1 c_1$ for the monopole and $J_0/3$ for the dipoles.

When a boundary or an interface is present, the reflected (or, generally speaking, backscattered) field changes the energy output of the source. The expression for the total output has been found in [A2, Sec. 8.7], and in the problem at hand can be written as follows:

$$J_t = \frac{J_0}{4\pi k_1} \int_{q<k_2} d^2\mathbf{q} \left[ |S_2(\mathbf{q})|^2 \exp(-2z_0 \operatorname{Im}\nu_1) \operatorname{Re}\left(\frac{1-|V|^2 + 2i\operatorname{Im}V}{\nu_1}\right) \right.$$
$$\left. + \left|S_1(\mathbf{q}) + V(q)S_2(\mathbf{q})e^{2i\nu_1 z_0}\right|^2 \operatorname{Re}\nu_1/\nu_1^2 \right]. \tag{A1}$$

Equation (A1) can also be derived directly by using the integral representation, eqs. (1) – (3), of the acoustic pressure and integrating the vertical component of the acoustic power flux over planes $z = z_0(1 \pm \varepsilon)$, $0 < \varepsilon \ll 1$, much in the same way that eq. (5) has been obtained.

Equations (5) and (A1) are valid for reflecting surfaces with arbitrary plane-wave reflection coefficients $V(\mathbf{q})$. Consider a monopole source of sound under a plane water-air interface. Using an explicit expression, eq. (4), for the Fresnel reflection coefficient, for the acoustic power flux into the air and the total emitted power flux we obtain:

$$J_a = mJ_0 \left[ A_1(n,m) + A_2(n,m,k_1 z_0) \right], \tag{A2}$$

$$J_t = J_0 \left[ T_1(n,m,kz_0) + T_2(n,m,k_1 z_0) \right]. \tag{A3}$$

Here $A_1$ and $T_1$ describe contributions of homogenous plane waves in the water, while $A_2$ and $T_2$ describe contributions of those inhomogeneous plane waves in the water that become homogeneous plane waves in the air after refraction at the water/air interface [A3]. Dimensionless quantities $A_{1,2}$ and $T_1$ are given by the following integrals:

$$A_1(n,m) = \int_0^1 \frac{\sqrt{n^2 - u}\, du}{\left(m\sqrt{1-u} + \sqrt{n^2 - u}\right)^2}, \tag{A4}$$

$$A_2(n,m,b) = \int_1^{n^2} \frac{\sqrt{n^2 - u}\, \exp\left(-2b\sqrt{u-1}\right) du}{n^2 - m^2 - (1-m^2)u}, \quad b \geq 0, \tag{A5}$$

$$T_1(n,m,b) = 1 + \frac{1}{2} \int_0^1 \frac{du}{\sqrt{1-u}} \frac{m\sqrt{1-u} - \sqrt{n^2-u}}{m\sqrt{1-u} + \sqrt{n^2-u}} \cos\left(2b\sqrt{1-u}\right). \tag{A6}$$

The remaining quantity, $T_2$, is related to $A_2$:

$$T_2(n,m,b) = mA_2(n,m,b). \tag{A7}$$

As long as the mass density ratio $m \ll 1$, it is sufficient to calculate $A_{1,2}(n, m = 0)$ and the two first terms of $T_1$ development in powers of $m$. From eqs. (A4) and (A5) we obtain

$$A_1(n,0) = \frac{2}{n+\sqrt{n^2-1}}, \quad A_2(n,0,b) = \pi\sqrt{n^2-1}\left[L_{-1}\left(2b\sqrt{n^2-1}\right) - I_1\left(2b\sqrt{n^2-1}\right)\right], \quad (A8)$$

where $L_j$ and $I_j$ are modified Struve and modified Bessel functions (see, e.g., [A4, Chaps. 9 and 12]). Note that $A_2(n, m, b)$ is a monotone decreasing function of $b$. At $b \gg 1$,

$A_2(n,m,b) = 0.5b^{-2}(n^2-1)^{-1/2}\left[1+O(b^{-2})\right]$. When $b$ increases from zero to infinity, $A_2(n, 0, b)$ decreases from $2\sqrt{n^2-1}$ to zero.

From eq. (A6) we find

$$T_1(n,m,b) = 1 - \frac{\sin 2b}{2b} + 2m\int_0^1 \frac{u\,du}{\sqrt{n^2-1+u^2}}\cos(2bu) + O(m^2). \quad (A9)$$

In the sum $T_1(n,m,kz_0) + T_2(n,m,kz_0)$ in eq. (A3), the term $1 - (2k_1z_0)^{-1}\sin(2k_1z_0)$ dominates unless $kz_0 \ll 1$. Corrections $O(m)$ become significant when $k_1z_0 \sim (mn)^{1/2}$ or smaller. Numerical integration in eq. (A9) can be avoided when $n^2 \gg 1$ by developing the radicand in powers of $(u^2-1)/n^2$. In particular, retaining the two first terms of the development, one obtains

$$T_1(n,m,b) = 1 - \frac{\sin 2b}{2b} + \frac{2m}{n}\left(\frac{\sin 2b}{2b} - \frac{\sin^2 b}{2b^2}\right) + O\left(\frac{m}{n^3} + m^2\right). \quad (A10)$$

The integrals in eqs. (A6) and (A9) cannot be evaluated analytically in a closed form when $b$ is arbitrary but are easily solved when $b = 0$. From eqs. (A3), (A5), and (A6) we obtain the total source output for a very shallow source:

$$J_t\big|_{z_0=+0} = J_0 T_3(n,m), \tag{A11}$$

$$T_3(n,m) = \frac{2m\sqrt{n^2-1}}{1-m^2}$$
$$\times \left[ \frac{n-m}{\sqrt{n^2-1}} - \frac{\pi m}{2\sqrt{1-m^2}} + \frac{m}{\sqrt{1-m^2}} \left( 2\arctan\frac{n+m-\sqrt{n^2-1}}{\sqrt{1-m^2}} - \arctan\frac{m}{\sqrt{1-m^2}} \right) \right] \tag{A12}$$

and bounds for the total source output at an arbitrary depth:

$$0 < J_t/J_0 \leq 1 - (2k_1 z_0)^{-1} \sin(2k_1 z_0) + T_3(n,m). \tag{A13}$$

Equations (A11) – (A13) are exact and apply when $m < 1$ and $n \geq 1$.

Power output of a point monopole source (such as a source of volume velocity, see [A2, Sec. 4.1]), according to eq. (A1), depends on the reflective properties of and distance to the interface. In the limit $m \to 0$, we have a pressure-release boundary, for which $J_t = 1 - (2k_1 z_0)^{-1} \sin(2k_1 z_0)$. $J_t$ deviates from unity because of the interference of fields of the actual source and the image source at $(0, 0, -z_0)$. When a monopole source approaches a pressure-release boundary and the volume velocity produced by the source is kept constant, the energy output vanishes because fields due to the actual source and its image interfere destructively. Unlike this ideal case, the destructive interference is never complete for the water-air interface. According to eqs. (A3), (A6), and (A7), the energy output of a shallow source equals

$$J_t/J_0 = 2mn - \pi m(n^2-1)k_1 z_0 + \frac{2}{3}\left[1 + 2mn(2n^2-3)\right]k_1^2 z_0^2 + O\left(m^2 + k_1^3 z_0^3(m + k_1 z_0)\right) \quad (A14)$$

and has a positive minimum at

$$k_1 z_0^{(\min)} \approx \frac{3\pi(n^2-1)m}{4\left[1 + 2mn(2n^2-3)\right]} \quad (A15)$$

(see Fig. 1D). The minimum value of $J_t$ is about 0.01 of the power output of the same source when located far away from the interface.

Of the two contributions in eq. (A2) to the acoustic energy transmitted through the interface, the first one, which is proportional to $A_1$ and is due to homogeneous plane waves, coincides with the result that would be obtained in the ray theory. In the wave theory, there is an additional contribution due to inhomogeneous waves. Its significance depends on the source depth and the refraction index (Fig. 3A) and is insensitive to the mass density ratio as long as the latter is small. The contribution of homogeneous waves is depth-independent, while the contribution of inhomogeneous waves, quite naturally, decreases with increasing source depth. Note that, according to the analytical results presented above, $A_2$ increases with $n$ at small $k_1 z_0$ and decreases with $n$ at large $k_1 z_0$. With relatively large values of $n$ characteristic of the water-air interface, the contribution of inhomogeneous waves exceeds the contribution of homogeneous waves by the factor $\sqrt{n^2-1}\left(n + \sqrt{n^2-1}\right)$, or almost 16 dB when $n = 4.5$, for shallow sound sources (Fig. 3A). Put differently, the wave theory predicts $J_a/J_0 \approx 2mn$ for shallow sources instead of the much lower prediction $J_a/J_0 \approx m/n$ of the ray theory.

As the source depth increases, the share of the energy radiated into the air monotonically and rapidly decreases because of the decreasing role of inhomogeneous waves and, for depths greater than $z_0^{(\min)}$ defined in eq. (A15), also because of an increase in the total energy output. At $k_1 z_0 > 1$, $J_a$ is close to the value predicted within the ray theory; the remaining depth dependence of the ratio $J_a / J_t$ in Fig. 4B is mostly due to depth dependence of $J_t$. At $k_1 z_0 < 0.21$, at least 10% of the radiated energy is transmitted into the air. For orientation, at $f = 1$ Hz, this source depth range approximately corresponds to $z_0 < 50$ m.

Details of calculations for other point sources are similar and will not be presented here. As far as a wave theory of sound transmission through an interface is concerned, the main qualitative distinction of dipole and higher multipole sources from the monopole source lies in a greater weight of inhomogeneous waves in plane-wave spectra of multipoles. For underwater sources, it results in an even stronger increase in the acoustic power $J_a$ transmitted by a shallow multipole into the air, as compared to the power transmitted from the same source when it is located at a depth of a few wavelengths. In particular, for the horizontal dipole (where $S_1 = S_2 = iq_1/k_1$), $J_a$ increases by the factor $2n^3 / \left[ 2n^3 - (2n^2 + 1)(n^2 - 1)^{1/2} \right]$ (or by 30.3 dB if $n = 4.5$, cf. Fig. 3B) when $k_1 z_0$ decreases from infinity to zero. The analogous amplification factor for the vertical dipole (where $S_1 = -S_2 = iv_1/k_1$) is $\left[ 4(n^2 - 1)^{3/2} - 2n^3 + 3n \right] / \left[ 2(n^2 - 1)^{3/2} - 2n^3 + 3n \right]$, or by 30.0 dB if $n = 4.5$, cf. Fig. 3C.

## *Acoustic transparency of a rough interface*

The wavelength of low-frequency acoustic waves, such as infrasound, in air and, perforce, in water, is much greater than height of waves on the ocean surface. Therefore, the effect of surface roughness on the transmission of infrasound is generally expected to be small. However, with the acoustic power flux through the plane water-air interface being proportional to the small mass density ratio *m*, even small (compared to $J_0$) corrections due to roughness may be significant unless the power flux in the scattered waves is also proportional to the small parameter *m*.

We now consider the water-air interface as a time-independent and stationary (in the statistical sense) *random* surface. Random surface elevations $\eta(\mathbf{r})$ have zero mean, variance $\sigma^2$, and power spectrum $F(\mathbf{q})$. In terms of the spectrum $\tilde{\eta}(\mathbf{q})$ of realizations of the random surface, these properties can be written as follows:

$$\langle \tilde{\eta}(\mathbf{q}) \rangle = 0, \quad \langle \tilde{\eta}(\mathbf{q}) \tilde{\eta}^*(\mathbf{q}') \rangle = \sigma^2 F(\mathbf{q}) \delta(\mathbf{q} - \mathbf{q}'), \tag{A16}$$

where angular brackets denote statistical average and $\delta$ is Dirac's $\delta$-function. Assuming that rms slopes $\tau$ of the random surface are small compared to unity and roughness height is small compared to the acoustic wavelength and the source depth:

$$\tau \ll 1, \quad \sigma \ll \min\left(k_2^{-1}, z_0\right), \tag{A17}$$

we will use the method of small perturbations. For our purposes it is sufficient to calculate the main term of the development of the energy flux through the random surface in powers of the small parameters *m* and $k_1\sigma$.

Scattering of sound at a slightly uneven interface of two fluids has been considered by Bass and Fuks [A5], Voronovich [A6], and others. When a plane wave $p = \exp(i\mathbf{q}_0 \cdot \mathbf{r} - i\nu_{10}z)$, where $\nu_{10} \equiv \nu_1(\mathbf{q}_0)$, is incident on the rough surface from the water, the average reflected field is again a plane wave with a reflection coefficient $\overline{V}(\mathbf{q}_0)$ that depends on the parameters of roughness. The mean (coherent) reflection coefficient was calculated by Voronovich [A6] in the second approximation of the method of small perturbations and, to the leading order in $\sigma$, is given by the following equations (see [A2, p. 109]):

$$\overline{V}(\mathbf{q}_0) = V(\mathbf{q}_0) - \frac{2\nu_{10}\sigma^2}{(m\nu_{10} + \nu_{20})^2}\left\{mD_{11} - \nu_{20}^2(m-1)^2 D_{22} + m\nu_{20}\left[D_{12} + k_1^2(n^2 - 1)\right]\right\}, \quad (A18)$$

$$D_{js}(\mathbf{q}_0) = \int d\mathbf{q}\, F(\mathbf{q} - \mathbf{q}_0) d_{js}(\mathbf{q}_0, \mathbf{q})/(m\nu_1 + \nu_2), \quad j, s = 1, 2, \qquad (A19)$$

$$d_{11}(\mathbf{q}_0, \mathbf{q}) = \left[k_1^2(m - n^2) - (m-1)\mathbf{q}_0 \cdot \mathbf{q}\right]^2, \quad \nu_{20} \equiv \nu_2(\mathbf{q}_0),$$
$$d_{12}(\mathbf{q}_0, \mathbf{q}) = 2(\nu_1 + \nu_2)\left[k_1^2(m - n^2) - (m-1)\mathbf{q}_0 \cdot \mathbf{q}\right], \quad d_{22}(\mathbf{q}_0, \mathbf{q}) = \nu_1\nu_2. \qquad (A20)$$

The leading order in $k_1\sigma$ of the asymptotics of the incoherent component $p_s$ of the scattered field can be calculated in the first approximation of the method of small perturbations. In water, the leading term of the asymptotics is given by (see [A2, p. 113]):

$$p_s(\mathbf{R}) = \int d\mathbf{q}\, B(\mathbf{q}_0, \mathbf{q})\tilde{\eta}(\mathbf{q} - \mathbf{q}_0)\exp(i\mathbf{q} \cdot \mathbf{r} + i\nu_1 z), \qquad (A21)$$

where

$$B(\mathbf{q}_0, \mathbf{q}) = \left[1 + V(q_0)\right]\frac{i(1-m)}{m\nu_1 + \nu_2}\left[\frac{\nu_2\nu_{20}}{m} + k_1^2\frac{m - n^2}{m - 1} - \mathbf{q}_0 \cdot \mathbf{q}\right]. \qquad (A22)$$

For definitiveness, consider an omni-directional, monopole point sound source. Then the incoherent component of the scattered field is obtained by using the incident spherical wave decomposition into plane waves, eqs. (1) and (2) with $S_1 = S_2 = 1$, with subsequent integration of the scattered field, eq. (A21), caused by each incident plane wave over $\mathbf{q}_0$. The coherent component of the scattered field in water is given by eqs. (1) and (3) with $S_1 = S_2 = 1$, where the mean reflection coefficient, eq. (A18), should be substituted for $V$.

As long as $\langle p_s \rangle = 0$, the coherent and incoherent components of the scattered field give additive contributions to the mean acoustic power flux. Using eqs. (A16) and (A21), for the acoustic power flux into the air due to the incoherent scattered field in the water, we obtain

$$J_{sc} = -\int\int_{-\infty}^{+\infty} dxdy \, \frac{1}{\omega \rho_1} \text{Im} \left\langle p_s^* \frac{\partial p_s}{\partial z} \right\rangle \Bigg|_{z=+0}$$
$$= \frac{-\sigma^2}{\omega \rho_1} \int \frac{d\mathbf{q}_0}{|v_{10}^2|} e^{-2z_0 \text{Im} v_{10}} \int_{q<k} d\mathbf{q} \, v_1 |B(\mathbf{q}_0, \mathbf{q})|^2 F(\mathbf{q} - \mathbf{q}_0). \quad (A23)$$

We took into account that the power spectrum $F$ of roughness is a real-valued function.

The acoustic power flux into the air due to the coherent component of the field is given by eq. (5), provided that the mean reflection coefficient, eq. (A18), is substituted for $V$. Since $\bar{V} - V = O(k_1^2 \sigma^2)$, the difference between the mean scattered field and the field unperturbed by roughness accounts for an $O(k_1^2 \sigma^2)$ addition, $J_{av}$, to the power flux $J_a$ (see eq. (5)) through the plane interface. Neglecting terms $O(k_1^4 \sigma^4)$, we can find $J_{av}$, from eqs. (5) and (A18):

$$J_{av} = \frac{2}{\omega \rho_1} \left\{ -\int_{q_0<k_1} \text{Re}\left[\bar{V}(\mathbf{q}_0) - V(q_0)\right] \frac{V(q_0) d^2\mathbf{q}_0}{v_{10}} \right.$$
$$\left. + \int_{q_0>k_1} \exp(-2z_0 \text{Im}\, v_{10}) \text{Im}\left[\bar{V}(\mathbf{q}_0) - V(q_0)\right] \frac{d^2\mathbf{q}_0}{\text{Im}\, v_{10}} \right\}. \quad (A24)$$

Note that corrections to $J_a$ due to contributions of coherent and incoherent components of the scattered field are of the same (second) order in the small parameter $k_1 \sigma$.

Equations (A23) and (A24) apply to scattering of sound at an interface of arbitrary fluids. We need to evaluate $J_{sc}$ and $J_{av}$ only for $m \ll 1$. According to eq. (A22), $B(\mathbf{q}_0, \mathbf{q}) = 2i v_{10}$ in the limit $m \to 0$. Then, from eq. (A23) we obtain

$$J_{sc} = -\left[1 + O\left(m + k_1^2 \sigma^2 + \frac{\sigma^2}{z_0^2}\right)\right] \frac{4\sigma^2}{\omega \rho_1} \int d^2\mathbf{q}_0 e^{-2z_0 \text{Im}\, v_{10}} \int_{q<k_1} d^2\mathbf{q}\, v_1\, F(\mathbf{q} - \mathbf{q}_0). \quad (A25)$$

One needs to exercise certain care in evaluating the limit of the integrals in eq. (A24) at $m \to 0$. According to eq. (A18), $\bar{V} - V$ is proportional to $(m v_{10} + v_{20})^{-2}$. One cannot go to the limit $(m v_{10} + v_{20})^{-2} \to v_{20}^{-2}$ in the integrands, because that would result in unintegrable singularity at $q_0 = k_2$. However, it is easy to show that

$$\int f(\mathbf{q}_0) (m v_{10} + v_{20})^{-2} d^2\mathbf{q}_0 = O(\ln m) \quad (A26)$$

for any smooth, bounded function $f$ that tends to zero as $q_0^{-\alpha}$, $\alpha > 0$ or faster at $q_0 \to \infty$. Using eq. (A26), we obtain from eqs. (A24) and (A18) – (A20)

$$J_{av} = \left[1 + O\left(m \ln m + k_1^2 \sigma^2 + \frac{\sigma^2}{z_0^2}\right)\right] \frac{4\sigma^2}{\omega \rho_1} \int \text{Re}\left[D_{22}(\mathbf{q}_0)\big|_{m=0}\right] \exp(-2z_0 \text{Im}\, v_{10}) d^2\mathbf{q}_0$$
$$= \left[1 + O\left(m \ln m + k_1^2 \sigma^2 + \frac{\sigma^2}{z_0^2}\right)\right] \frac{4\sigma^2}{\omega \rho_1} \int d^2\mathbf{q}_0 e^{-2z_0 \text{Im}\, v_{10}} \int_{q<k} d^2\mathbf{q}\, v_1\, F(\mathbf{q} - \mathbf{q}_0). \quad (A27)$$

Note that, under the assumptions we made, both eqs. (A25) and (A27) contain a single, integral characteristic of the surface roughness; namely,

$$\sigma^2 \, \text{Re}\left[D_{22}(\mathbf{q}_0)\big|_{m=0}\right] = \sigma^2 \int_{q<k} d^2\mathbf{q} \, v_1 \, F(\mathbf{q}-\mathbf{q}_0). \tag{A28}$$

We see that both $J_{sc}$ and $J_{av}$ have non-zero limits when $m \to 0$. However, the leading terms of expansions of $J_{sc}$ and $J_{av}$ in powers of the small parameters of the problem cancel each other, according to eqs. (A25) and (A27). That is, within the approximation we consider, the total correction due to roughness to the acoustic power flux into the air equals zero.

Within the terms of the second order in $k_1\sigma$, expansion of $J_{sc} + J_{av}$ with respect to the small parameter $m$ starts from terms $m \ln m$ or smaller, and correction due to roughness is small compared to the acoustic power flux through a plane water-air interface as long as the condition eq. (A17) is met. The above analysis does not allow us to evaluate terms of higher order than $k_1^2\sigma^2$ in the expansion of $J_{sc} + J_{av}$. Still, these higher-order terms are expected to vanish in the limit $m \to 0$ since $J_{sc} + J_{av} = 0$ to any order in $k_1\sigma$ in the case of a rigid boundary, i.e., at $m = 0$.

Aside from its effect on energy flux into the air, surface roughness affects distribution of the acoustic energy between the water and the air by changing the source output. From the general expression for source output obtained in [A2, Sec. 8.7], it follows that perturbation in the energy output of a point monochromatic source of volume velocity is bilinear in source strength and in the acoustic pressure of the scattered wave at the point where the source is located. (We assume that the source strength, i.e., amplitude and phase of the monochromatic oscillations of volume velocity the source creates, is kept constant. In other words, we assume that the source radiates the same spherical wave $p_0$ whether or not the ocean surface is rough.) Hence, roughness-induced change in the statistical average of the source output depends only on the average scattered field. Consequently, in the case of a rough surface, the total energy output $J_t$ can still be calculated with eq. (A1) provided the Fresnel reflection coefficient $V$ is replaced by the mean reflection coefficient $\bar{V}$, eq. (A18).

Let $J_t = J_t^{(0)} + J_t^{(av)}$, where $J_t^{(0)}$ is the power output, eq. (A3), in the case of the plane interface and $J_t^{(av)}$ is a correction due to roughness. From eqs. (A1) and (A18) we obtain (cf. eq. (A27))

$$J_t^{(av)} = \left[1 + O\left(m \ln m + k_1^2 \sigma^2 + \frac{\sigma^2}{z_0^2}\right)\right] \frac{4\sigma^2}{\omega \rho_1} \operatorname{Re} \int d^2\mathbf{q}_0 e^{2i v_{10} z_0} \int d^2\mathbf{q} \, v_1 \, F(\mathbf{q} - \mathbf{q}_0). \quad (A29)$$

Being proportional to $\sigma^2$, the roughness-induced correction to the source output is always small compared to $J_0$. Hence, the correction can be significant only where $k_1 z_0 \ll 1$. When the latter inequality holds, following [A2, Sec. 3.6], in estimating integrals on the right side of eq. (A29), one obtains $J_t^{(av)} = k_1^2 z_0^2 J_0 O\left(\sigma^2/z_0^2 + \tau \sigma/z_0\right)$, where $\tau$ is the rms slope of the rough interface. Under validity conditions eq. (A17) of the method of small perturbations, the expression within the brackets is small compared to unity. Therefore, $J_t^{(av)} \ll J_t^{(0)}$ even where $k_1 z_0 \ll 1$. We see that corrections due to roughness to the source output and transparency of the interface are negligible as long as condition (A17) is met.